\documentclass[11pt]{article}
\usepackage{moriond,epsfig}

\def\be{\begin{equation}}
\def\ee{\end{equation}}
\def\bea{\begin{eqnarray}}
\def\eea{\end{eqnarray}}

\begin{document}
\vspace*{4cm}
\title{Signatures of Non-Standard Electroweak Symmetry Breaking}

\author{ Veronica Sanz\footnote{On leave of absence from Department of Physics and Astronomy, York University, Toronto, Canada. } }

\address{CERN, Theory Division, CH-1211, Geneva 23, Switzerland
}

\maketitle\abstracts{
.}

\section{Introduction: What is {\it  \bf Standard}?}
 
 This talk focuses on non-standard scenarios of electroweak symmetry breaking (EWSB). In this talk, {\bf Standard } means that EWSB  is triggered by a sector of elementary scalars. This includes the Standard Model (SM) and extended Higgs sectors such as in Supersymmetry~\cite{SUSY}.  
 
 Non-standard means anything else. Whether EWSB is triggered by a new sector of composite scalars, or no scalars at all, that is the meaning of non-standard in this talk.
Non-standard EWSB is realized in Extra-dimensions~\cite{xdims}, Little Higgs~\cite{LH}, and Technicolor~\cite{TC}. In those cases, the Higgs is either absent or composite~\cite{composite}.
 
 \section{Composite Higgs}
 
There is nowadays an increasing suspicion that there may be a light scalar particle at around 125 GeV in mass~\cite{Higgs125}~\footnote{Ok, I was supposed to start writing these proceedings two weeks ago, so bear with the 'strong suspicion'.}. A  {\it light} scalar needs a stabilization mechanism, or we {\it really} do not understand the formulation of scalar theories in quantum field theory. This is the hierarchy problem: why would the scalar stick around the electroweak scale, when any threshold correction from UV physics pushes its mass to UV values? 
 
  \vspace{.3cm}
  
 The usual answer is {\it symmetries, symmetries, symmetries!} 
 
 \vspace{.3cm}
 
If one opts for for fermionic symmetries, the name of the game is Supersymmetry. The Higgs is protected from UV contributions, hence its mass stabilized, by embedding the Higgs in a superfiled along with fermions. The Higgs inherits the protection of  chiral symmetry, if Supersymmetry is unbroken. Supersymmetry is obviously broken, and that makes the nice, cute story a whole mess, but Nature is sometimes funny. 

If one opts for bosonic symmetries, the names we have in mind are Nambu and Goldstone. Light scalar degrees of freedom may be just the manifestation of the spontaneous breaking of some approximate global symmetry. Pseudo-Goldstone bosons stick around as their mass is purely linked to IR effects, hence stable under radiative corrections. In the limit when all explicit sources of symmetry breaking are off, the scalar is massless.  

Both Supersymmetry and Pseudo-Goldstone Higgs have their beauty and their problems. And none, if there, manifests itself in a simple/minimal fashion. Those scenarios need gymnastics, model-building, tuning$\ldots$ creativity, in a word, to pass the stringent electroweak precision tests (and the always tightening collider bounds).

Supersymmetry has been by far the most popular option to stabilize the electroweak scale, and for many years dominated the spectrum of Beyond the Standard Model. And for a good reason: Supersymmetry is elegant, perturbative and, most importantly, easily tuneable to look like the Standard Model as the experimental results push the supersymmetric phenomena to higher and higher scales. 

But now that the experimental bounds are {\it really } getting the community nervous, there is an increasing attention to the view of the Higgs as a pseudo-Goldstone boson. In this case, the Higgs would be the equivalent of the pions and kaon sector in QCD, a composite of some IR confining theory, whose mass is much below the masses of other composite states because it is protected by an approximate shift symmetry.

\subsection{Realizations of Composite Higgs}

Composite Higgs can be realized in Little Higgs, Extra-dimensions and Technicolor. 

In Little Higgs, the Higgs is a pseudo-Goldstone of a new global symmetry. Well, actually, in Little Higgs we are talking about a new sector of global symmetries, which breaks down and are also partially gauged. 

In Extra-dimensions, the Higgs may be the extra-dimensional component of a gauge field in more than four dimensions. After compactification of the extra dimensions down to four, the compactification procedure may respect a subset of the original higher-dimensional gauge symmetries. This subset may include a shift symmetry, parallel to the Goldstone boson protection mechanism. In the AdS/CFT interpretation of extra-dimensions as dual to composite sectors, the compactification in extra-dimensions which leads to a remaining shift symmetry is dual to the Nambu-Goldstone mechanism.
The Higgs is then a holographic pseudo-Goldstone boson.

In Technicolor, there is no need for a scalar degree of freedom to break the electroweak symmetry. EWSB is driven by some new gauge sector. This sector becomes strong and triggers the breaking of symmetries, global and partially gauged, including the electroweak symmetry of the SM. Our known and beloved $W$ and $Z$ are themselves composites of, for example, the techni-fermions  charged under the new gauge symmetry. In this picture, $W$ and $Z$ eat would-be Goldstone bosons.

Although there is no need for a Higgs-like particle, technicolor dynamics can accommodate it. The pion sector may be quite complex, with some pions would-be goldstones, and some pseudo-goldstones which remain in the spectrum. 

\subsection{Symmetries for a light Higgs}

There are many examples of 4D symmetry breaking patterns resulting in three would-be goldstones and some light scalars. The literature on this is very long, but this long list is cut to few breaking patterns when one ask for one and only one light scalar.

Extra-dimensional gauge theories provide another UV completion of the Goldstone symmetry. In that case, it's a relic  shift symmetry from compactifying a gauge symmetry. After compactification, a higher dimensional gauge theory could break down to a 4D gauge theory or some relic shift symmetry, restricting the properties of the extra-dimensional component of the gauge field. 

\subsection{Generic Features of Composite Higgs}

What is the composite Higgs supposed to do? A {\it must} is the $W$$W$ scattering unitarization. The composite Higgs does part of the job of unitarizing, and the rest is done by heavy resonances. So, if the Higgs is composite, one could try to measure the degree of non-unitarization in WW scattering with the Higgs exchange only. But accessing to the unitarization process  (i.e. measuring precisely the vector boson fusion channel) is an herculean task,  quite late in the LHC program. But one can take a detour: in composite Higgs one also expects some deviation of the Higgs couplings to fermions and vector bosons respect to the SM. So, instead of measuring the VBF channel, one could analyze the production and decay of Higgs to $\gamma\gamma$, $ZZ$, $WW$, $b\bar{b}$  aand $\tau^+\tau-$ and discover that those couplings do not correspond to the SM expectation. Those deviations, and also deviation in VBF, are proportional to $(v/f)^2$, where $f$ is the scale of strong interactions. As in SUSY, one can always decouple these effects to the price of a larger fine-tuning, and deform the theory to look too alike to the SM. Even $f~$ 500 GeV is a tough measurement~\cite{VBF}.
 
 In summary, composite Higgs' generic features are (may be) {\it too } SM-like.
 
 \subsection{Common features}
 
 If $f$ is about or larger than 500 GeV, one has a better shot at trying to find the resonances themselves. Usually the mass of the resonances, bound states of new strong interactions of confinement scale $f$ is given by
 \bea
 m_{\rho} = g_{\rho} f
 \eea 
 where $g_{\rho}$ is a dimensionless coupling which encodes how narrow/broad the resonance is.
Unless those resonances are very broad ($g_{\rho} \gg 1$), the resonances shouldn't be far away, and one can expect to produce them directly. Let us use their spin to classify the new particles:

\begin{enumerate}
  
  \item {\bf New Fermions: }  The way we call these new fermions depend on the model.  For example, in Little Higgs, we would name it $T$,  the top partner, and their role is to cancel quadratic contributions to the Higgs mass. In extra-dimensions, the same guy would be called, $t_{KK}$, the Kaluza-Klein excitation of the top. In the extra-dimensional case we expect many, not just one, top partners. But since the first KK resonance is expected to be heavy, searching for the next resonance is quite a challenge. What about new strong interactions? If the top is (partly) composite, i.e. either it is a bound state, or an elementary fermion with some mixing with a bound state, we expect heavy technibaryons with the same quantum numbers.  
  
  In all these cases, the new fermion couples/mixes with SM fermions, hence it can decay to them. In that sense, the searches for new fermions are no different (at the most basic level) than searches for fourth generation fermions~\cite{fourth-gen}, and their interpretation would only be clarified if other resonances, and modified Higgs couplings were measured.
  
  \item {\bf New Vectors:} Again, the names and interpretation depend on the scenario (for example, $W'$, $W_{KK}$ and $\rho_{TC}$) but the phenomenology is very similar. Those resonances can be produced s-channel via their couplings to light fermions or, if those couplings are very suppressed, they can be produced in vector boson fusion. 
  
  The searches for spin-one resonances in these scenarios are no different that $Z'$ and $W'$ searches, so their discovery (alone) would be hard to disentangle from a new gauge symmetry, broken at the TeV scale.
  
  \item {\bf Massive spin-two resonances: } Now, this is an interesting, difficult to mimic scenario. One usually thinks about this creature as a KK graviton, but one should also keep in mind that any strongly coupled sector would produce a zillion of resonances, and among those spin-two. QCD has quite some of them, for example, the $f_2$. In this talk we will discuss the phenomenology of those two types of spin-two resonances and ways to distinguish them. I can anticipate that this task  will be harder that one would imagine.
\end{enumerate}

\section{No Higgs}

Well, to tell you the truth, I was finishing this proceedings as the Higgs discovery was announced~\footnote{An event which caused to many of us both incredible excitement and getting a new PHD ({\it PHD=Post-Higgs Depression}).}, and I cannot force myself to write a section about no-Higgs now. If I had the strength, I would tell you about distinguishing different scenarios using leptons at the LHC~\cite{dileptons}. 

What can I say? Cool idea, not realized in Nature. Let us move on.

\section{How to unmask a graviton impostor}

In composite/extra-dimensional theories, spin-two resonances may be lying around at low scales. Can we distinguish them? or, in other words, how do we discover them?

One would naively say that if this guy comes from extra-dimensions, then it is very easy to spot. KK-gravitons inherit their couplings from the massless graviton, 
\bea
\frac{1}{\Lambda} \, G_{\mu\nu} \, T^{\mu\nu}
\eea
where $G$ is the graviton excitation (a 4D field) and $T^{\mu\nu}$ is the ordinary stress-tensor. But there is something peculiar about this coupling. It is not the way I just wrote it, because it actually looks as
\bea
\frac{c_{i} }{\Lambda} \,  G_{\mu\nu} \, T_i^{\mu\nu}
\eea
where $i$ is a label for species, namely the graviton could couple to different fields with a different coefficient. So, yes, the coupling is still to the {\it structure} contained in the stress tensor, but the graviton does not have to couple to the SM fields universally. The origin of the $c_i$ is localization in the extra-dimension. SM fields can be localized on a brane, or move in the extra-dimensions with different wavefunctions, and there is a lot of freedom on what they can do and how they end up overlapping with the graviton wavefunction. The $c_i$'s are then very model dependent.

But then a spin-two resonance from some unknown strongly coupled sector, let us call it the impostor $\tilde{G}$, would couple in a more generic way, right? The answer is {\bf no}. The reason, guess what? symmetries, symmetries, symmetries! The only thing we know how to do well.

At quadratic level in fields, the Lagrangian of any rank-2 tensor is the Pauli-Fierz. So they do propagate in the same way. And at the level of interactions, after imposing Lorentz and CP invariance, one can show~\cite{us} that all the interactions between the impostor and the SM look like
\bea
\frac{\tilde{c}_{i} }{\Lambda} \,  \tilde{G}_{\mu\nu} \, T_i^{\mu\nu}
\eea 
Does it sound familiar? The impostor couples to the same structures as the graviton. The interpretation of $\tilde{c}$ is different, though. The resonance would couple to SM fields which {\it talk} to the strong sector. The top, the massive $W$ and $Z$ are clear candidates. But the photon, or gluon could also couple with the same mechanism and the $\rho-\gamma$ mixing of QCD. 

Now, if you are tuned to holographic models, this comes as no surprise. At the end of the day, one could imagine that, for every sector of strong dynamics, there is {\it some} extra-dimensional dual, no matter how complicated its geometry. This duality can be exploited ad infinitum, see for example Refs.~\cite{analogue,sum-rules} for some fun I had with it.

Then, if you are a holographic lover, you are going to be surprised: the duality does not hold here. The photon and the gluon are massless, manifestation of the conservation of $U(1)_{EM}$ and $SU(3)_C$ above the confinement scale. Those are currents which have to be coupled in the extra-dimensional model in a consistent way. If photons and gluons propagate in the extra-dimension, then one can define a ratio of partial decay widths
\bea
R=\frac{\Gamma(\to g g )}{\Gamma(\to \gamma\gamma)}
\eea
which in extra-dimensions is fixed to 8, whereas in a composite sector it can be anything. Techni-fermions could be charged under color and/or EM, and the spin-two bound state could decay to photons and not to gluons, for example. In the paper we discuss how difficult is to measure this ratio, but it is an explicit example of the  breakdown of the correspondence.

\section*{Acknowledgments}
I thank my collaborators, obviously.


\begin{thebibliography}{99}
\bibitem{Higgs125} Where were you all this time, in a desert island with no internet connection?
\end{thebibliography}
\end{document}